\documentclass[twocolumn,prb,showpacs,floats]{revtex4}
\usepackage[dvips]{graphicx}
\usepackage{dcolumn}
\usepackage{color}

\usepackage{multirow} 

\usepackage{bm}

\newcommand{\SMIO}{Sr$_3$$M$IrO$_6$}
\newcommand{\SNIO}{Sr$_3$NiIrO$_6$}
\newcommand{\SCIO}{Sr$_3$CoIrO$_6$}

\begin{document}

\title{Impact of spin-orbit coupling on the magnetism of {\SMIO} ($M$ = Ni, Co)} 
\author{Xuedong Ou}
\affiliation{Laboratory for Computational Physical Sciences (MOE), State Key Laboratory of Surface Physics, and Department of Physics, Fudan University, Shanghai 200433, China}
\author{Hua Wu}
\thanks{Corresponding author. wuh@fudan.edu.cn}
\affiliation{Laboratory for Computational Physical Sciences (MOE), State Key Laboratory of Surface Physics, and Department of Physics, Fudan University, Shanghai 200433, China}

\date{today}

\begin{abstract}

Using density functional calculations, we demonstrate that the spin-orbit 
coupling (SOC) of the Ir$^{4+}$ ion plays an essential role in determining 
the antiferromagnetism of the hexagonal spin-chain system {\SMIO} ($M$ = Ni, Co) 
by tuning the crystal-field level sequence and altering the Ir-$M$ 
inter-orbital interactions. The SOC splits the $e'_g$ doublet of 
the octahedral Ir$^{4+}$ ion ($t_{2g}^5$) in a trigonal crystal field, and the 
single $t_{2g}$ hole resides on the $e'_g$ upper branch and 
gives rise to the antiferromagnetic superexchange.
In absence of the SOC, however, the single $t_{2g}$ hole would occupy 
the $a_{1g}$ singlet instead, which would mediate an unreal ferromagnetic exchange 
due to a direct $a_{1g}$ hopping along the Ir-$M$ chain. 
We also find that the Ni$^{2+}$ and Co$^{2+}$ ions are both in a high-spin state
and moreover the Co$^{2+}$ ion carries a huge orbital moment.
This work well accounts for the recent experiments and magnifies again 
the significance of the SOC in iridates. 

\end{abstract}

\pacs{75.25.Dk, 71.20.-b, 71.70.-d}

\maketitle

Charge, spin and orbital states are often coupled in $3d$ transition-metal
oxides due to their multiple degrees of freedom and electron correlation.
These states are closely related to diverse material properties and 
functionalities, e.g., charge ordering, orbital ordering, spin-state and 
magnetic transition, metal-insulator transition, superconductivity, 
colossal magnetoresistance,
and multiferroicity. It is therefore very important to study those 
charge-spin-orbital states and their fascinating coupling for modeling and 
understanding of the abundant properties. This has formed a research stream
in condensed matter physics over past decades, see e.g., 
a short review~\cite{Dagotto}. Very recently, research 
interest has been extended to $5d$ transition-metal oxides, which probably 
possess a significant spin-orbit coupling and provide an avenue to novel
magnetic and electronic properties due to an entangled spin-orbital state.

In this respect, iridates are a representative example~\cite{Kim08,Kim09,
Shitade,Pesin,Wang,Jackeli,Chaloupka,Wan11,Yin,Haskel,Mazin,
Katukuri,Gretarsson}. An octahedrally
coordinated iridium ion normally has a large $t_{2g}$-$e_g$ crystal-field
splitting due to the delocalized character of its $5d$ electrons. The 
resultant low-spin state with only a $t_{2g}$ occupation makes an open-shell
Ir$^{n+}$ ion (e.g., $t_{2g}^5$ for $n$ = 4) behave effectively like 
$p$ electrons (with an effective orbital momentum {\it\~l} = 1). As a result,
an intrinsic strong spin-orbit coupling (SOC) splits the $t_{2g}$ levels 
into a lower
{\it\~j} = 3/2 quartet and a higher {\it\~j} = 1/2 doublet. Then, for an Ir$^{4+}$
constituent oxide, the half-filled {\it\~j} = 1/2 doublet may form, due to 
a moderate electron correlation, a novel {\it\~j} = 1/2 Mott 
insulating state~\cite{Kim08,Kim09}. 
It has been proposed that such a spin-orbital entangled state can bring about
exotic properties, e.g., correlated topological insulator~\cite{Shitade,Pesin},
superconductivity~\cite{Wang}, Kitaev model~\cite{Jackeli,Chaloupka},
Weyl semimetal~\cite{Wan11}, and unusual magnetism~\cite{Yin}.  

\begin{figure}[b]
\centering \includegraphics[width=8cm]{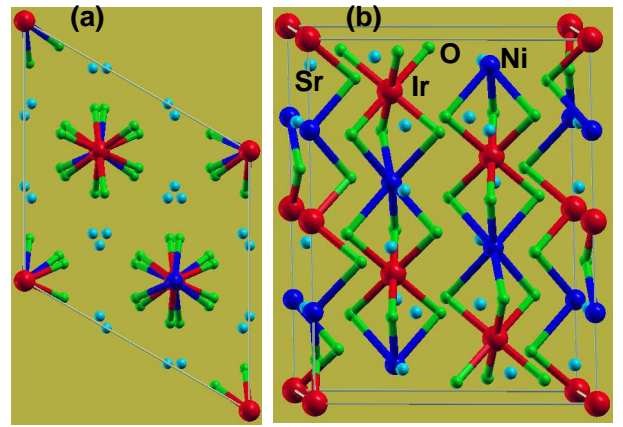} 
\caption{(Color online)
Crystal structure plot of {\SMIO} (a) projected onto the $ab$ plane and (b) 
in a perspective view. It has a hexagonal $ab$ plane and quasi one-dimensional
$M$IrO$_6$ spin chains extending along the $c$-axis, in which the IrO$_6$ 
octahedra and $M$O$_6$ trigonal prisms are alternating.
}
\end{figure}

In this paper, we have studied the $3d$-$5d$ transition-metal
hybrid material {\SMIO} ($M$ = Ni, Co), and find that the SOC has a significant
impact on its magnetism by tuning its spin-orbital states and the Ir-$M$
inter-orbital interactions. This system has a general chemical
formula $A_3$$MM'$O$_6$ ($A$ = Ca, Sr; $M$ = $3d$ TM, $M'$ = $3d$, $4d$, $5d$ 
TM) and displays an in-plane hexagonal structure and out-of-plane spin chains,
see Fig. 1. Those quasi one-dimensional spin chains each consist of 
alternating face-sharing $M$O$_6$ trigonal prisms and $M'$O$_6$ octahedra. 
This system drew a lot of attention in the past decade~\cite{Niitaka,Wu05,
Choi,Wu09,Agrestini,Kamiya}, because of its 
intriguing step-wise magnetization, significant Ising-like magnetism, 
thermoelectricity, and multiferrocity. {\SNIO} and {\SCIO}
also possess fascinating magnetism~\cite{Nguyen,Tjeng,Zhang,Sarkar}. 
Owing to their complex temperature-dependent magnetic transitions,
a standing issue is the nature of their dominant intrachain magnetism: either 
an intrachain ferromagnetic (FM) exchange~\cite{Nguyen,Sarkar} or an 
antiferromagnetic (AF) coupling~\cite{Tjeng,Zhang} was proposed in previous
studies. Moreover, the origin of the magnetism remains elusive.
Therefore, {\SNIO} and {\SCIO} call for a prompt study to clarify the nature and origin of their intriguing intrachain magnetism.
As seen below, we make a comparative study for {\SNIO} and {\SCIO}, by 
carrying out a systematic set of electronic structure calculations. Our results 
consistently explain the experimental observations and settle the standing issue. 
In particular, we find that the SOC of the Ir$^{4+}$ ion plays an essential role 
in determining the intrachain AF structure of {\SMIO} ($M$ = Ni, Co) 
by tuning the crystal-field 
level sequence and altering the Ir-$M$ inter-orbital interactions. 
Therefore, {\SMIO} is added to the
iridate category which highlights the significance of the SOC.


We have carried out density functional calculations, using the full-potential
augmented plane wave plus local orbital code (Wien2k)~\cite{Blaha}. 
We use the structural 
data of {\SNIO} measured by a neutron diffraction at 10 K~\cite{Nguyen} 
and of {\SCIO} at 4 K~\cite{Tjeng}. 
They have practically the same crystal structure: the Ir-O bondlength of 
2.01~\AA, the $M$-O 2.18~\AA, and the $M$-Ir 2.78~\AA~ for $M$ = Ni and Co;
the small deviation of the O-Ir-O bond angle (from the ideal 90$^{\circ}$) 
due to a small trigonal 
distortion of the IrO$_6$ octahedron, being 5.4$^{\circ}$ for $M$ = Ni 
and 5.2$^{\circ}$ for $M$ = Co.  
The muffin-tin sphere radii are chosen to be 
2.8, 2.2, 2.2 and 1.5 Bohr for Sr, Ni/Co, Ir and O atoms, respectively. 
The plane-wave cut-off energy of 16 Ry is set for the interstitial wave 
functions, and $5\times5\times5$ {\bf k} mesh for integration over the 
rhombohedral Brillouin zone. We employ the local spin density approximation 
plus Hubbard $U$ (LSDA+$U$) method~\cite{Anisimov} to describe the electron
correlation of the $M$ $3d$ and Ir $5d$ electrons. 
The typical values, effective $U$ = 2, 4 and 5 eV are used for the Ir $5d$, 
Co $3d$, and Ni $3d$ states, respectively. 
Note that our key results --- the coupled spin-orbital state and the 
magnetic ground state --- are independent of the tested $U$ values (1-3 eV for
Ir $5d$ and 3-7 eV for Co/Ni $3d$).
To account for (near) degeneracy of the Ir $5d$ orbitals (and of $M$ $3d$ 
orbitals as well), the SOC is included by the second-variational method with 
scalar relativistic wave functions. 
In order to probe diverse possible spin-orbital states and magnetic structures, 
we would excess them
in our calculations by setting their respective occupation number matrix
and thus orbitally dependent potentials, and then do self-consistent
calculations including a full electronic relaxation.
(Otherwise, some states of the concern or even a ground state cannot be achieved.)
An advantage of this procedure is such that we can reliably determine the magnetic ground state 
by a direct comparison of the different states.~\cite{Wu05,Wu09}

\begin{figure}[t]
\centering \includegraphics[width=8cm]{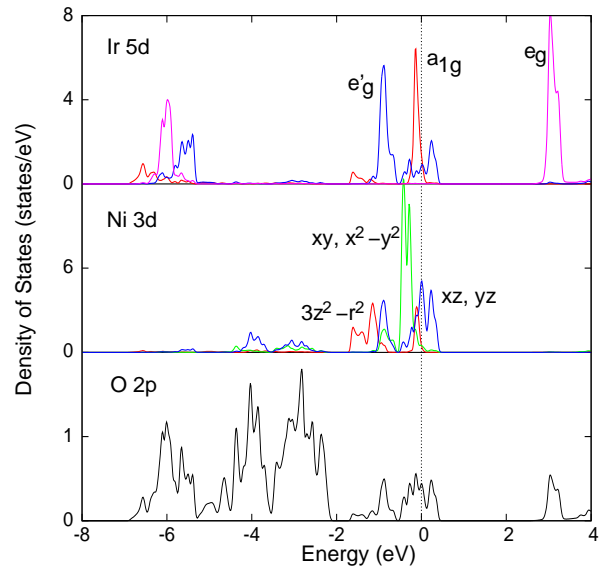} 
\caption{(Color online)
Partial density of states (DOS) of {\SNIO} in the nonmagnetic state 
calculated by LDA. The octahedral Ir ion has a common large 
$t_{2g}$-$e_g$ crystal
field splitting of more than 3 eV; and in a trigonal crystal field (elongation
of the IrO$_6$ octahedron along the local [111] direction, i.e., the
$z$-axis of the hexagonal lattice), 
the $t_{2g}$ splits further into
a lower $e'_g$ doublet and a higher $a_{1g}$ singlet.  
The trigonal prismatic coordination produces the Ni $3d$ crystal-field
level sequence (from low to high) as $3z^2$--$r^2$ / xy, $x^2$--$y^2$ / xz, yz.
}
\end{figure}

We first carry out spin-restricted LDA calculations to estimate the crystal
field splitting. The calculated DOS (density of states) results for {\SNIO}
are shown in Fig. 2. The O $2p$ valence bands lie in between --7 and --2 eV.
They have a significant covalency with the delocalized Ir $5d$ orbital and
bring about the Ir $5d$ bonding state around --6 eV. A relatively weak 
Ni-O hybridization yields the Ni $3d$ bonding state in between --2 and
--4 eV. Both the Ir $5d$ and Ni $3d$ antibonding states lie above --2 eV.
For the Ir ion, its local octahedral coordination but a trigonal crystal 
field in the global coordinate system split the otherwise $t_{2g}$ triplet 
into the $e'_g$ doublet and the $a_{1g}$ singlet both of the concern.
The $e_g$ doublet is far above them by 3 eV and is out of the concern.
The $a_{1g}$ singlet can be written as $3z^2$--$r^2$, as the $z$-axis
of the hexagonal lattice is along the [111] direction of the local IrO$_6$
octahedra. Moreover, the $e'_g$ doublet can be expressed as 
$\sqrt{2/3}$ $xy$ + $\sqrt{1/3}$ $yz$ and 
$\sqrt{2/3}$ ($x^2$--$y^2$) -- $\sqrt{1/3}$ $xz$, 
when the $y$-axis is set  
along the [1\=10] direction of the local IrO$_6$ octahedra (then the 
$x$-axis is uniquely defined and the $xy$ is in the hexagonal $ab$ plane).
By integrating the DOS and determining the center of gravity for each 
eigen orbital within the antibonding energy range, we find that the $e'_g$
doublet is lower than the $a_{1g}$ singlet by 0.21 eV. This is consistent
with the structural feature that the IrO$_6$ octahedron is slightly elongated 
along the hexagonal $z$-axis and the resultant O-Ir-O bond angle deviates 
from the ideal 90$^{\circ}$ by only 5.4$^{\circ}$~\cite{Nguyen}. 
As a result, the $a_{1g}$ has a higher
crystal-field level than the $e'_g$. For the Ni ion, its trigonal prismatic
coordination produces the crystal-field level sequence of the Ni $3d$ 
electrons as $3z^2$--$r^2$ / $xy$, $x^2$--$y^2$ / $xz$, $yz$ (0 / 0.57 / 0.76 eV). 
Furthermore, there are inter-site interactions between the Ir $5d$ and 
Ni $3d$ orbitals, see Fig. 2. The Ir $a_{1g}$ ($3z^2$--$r^2$) and Ni 
$3z^2$--$r^2$ electrons
have a lobe pointing to each other and have a $dd\sigma$ hybridization.
The Ir $e'_g$ orbital has, via its $xz$ or $yz$ component, a $dd\pi$ 
hybridization with the Ni $xz$/$yz$; and via its $xy$ or $x^2$--$y^2$ 
component, a weak $dd\delta$ hybridization with the Ni $xy$ / $x^2$--$y^2$.
Note that those crystal-field level sequences and the Ir-Ni inter-orbital 
hybridizations are crucial for understanding of the spin-orbital state and the 
intrachain magnetism in {\SMIO}, as seen below.

Then we perform spin-polarized LSDA calculations. We start with a FM or an AF 
state with the Ni$^{2+}$ spin=1 and Ir$^{4+}$ spin=1/2, but both calculations
converge to a same FM metallic solution (not shown here). It has a total spin
moment of 2.82 $\mu_B$/fu, consisting of the local spin moments of 
1.46 $\mu_B$/Ni$^{2+}$, 0.54 $\mu_B$/Ir$^{4+}$, and 0.11 $\mu_B$/O. The
Ni$^{2+}$ ion has the electronic configuration 
(3$z^2$--$r^2$)$^2$($xy$, $x^2$--$y^2$)$^4$($xz$, $yz$)$^2_{\uparrow}$,
and the Ir$^{4+}$ has a single $t_{2g}$ hole mostly on the $a_{1g}$ orbital
(i.e., ($e'_g$)$^4$$a_{1g\uparrow}^1$). The Ir-O and Ni-O covalencies
bring about an appreciable spin moment of 0.11 $\mu_B$ on each oxygen.
As the $a_{1g}$ orbital is a higher crystal-field level than the $e'_g$,
the $a_{1g}$ hole state allows a direct $a_{1g}$ electron hopping from
Ni$^{2+}$ to Ir$^{4+}$, and this prompts the FM metallic solution 
(cf., Figs. 4(a) and 4(b)). 

\begin{figure}[t]
\centering \includegraphics[width=8cm]{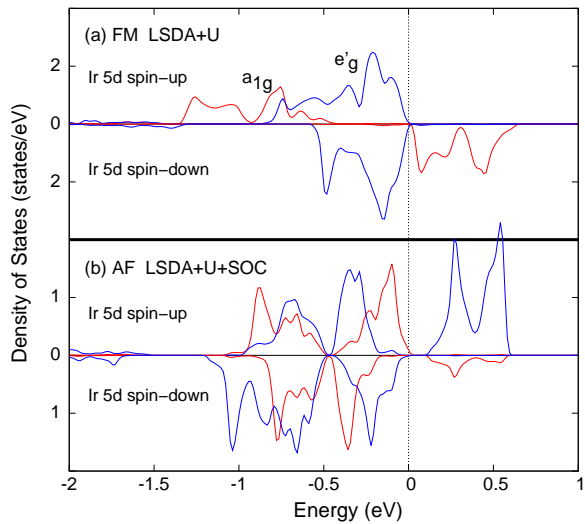} 
\caption{(Color online)
The partial DOS of the Ir-$5d$ $a_{1g}$ and $e'_g$ 
orbitals in {\SNIO} in (a) the FM state calculated by LSDA+$U$ and in (b)
the AF state by LSDA+$U$+SOC. Taking the Ni$^{2+}$ $S$=1 as a reference, the
Ir$^{4+}$ ion has a down-spin $a_{1g}$ empty state in (a) but an up-spin
$e'_g$ empty state (i.e., a complex orbital with $l_z$ = 1) in (b).
}
\end{figure}
 
\begin{figure}[b]
\centering \includegraphics[width=7cm]{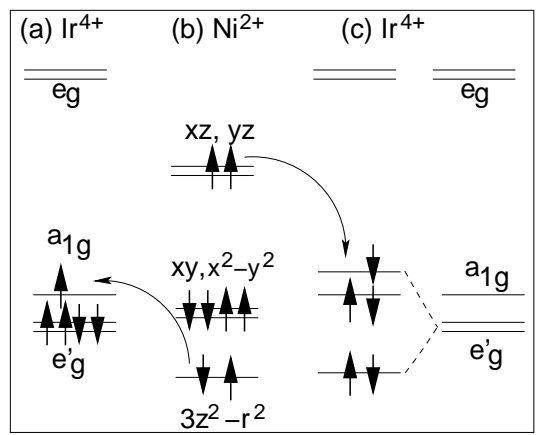} 
\caption{
Schematic level diagrams of the Ir$^{4+}$ $5d$ and
Ni$^{2+}$ $3d$ orbitals. The down-spin $3z^2$--$r^2$ electron mediates a FM
coupling via a $dd\sigma$ hybridization [(a) and (b)]. The up-spin $xz$/$yz$
electrons mediate an AF coupling via a $dd\pi$ hybridization
[(b) and (c)], in which the SOC splits the $e'_g$ doublet. 
}
\end{figure}

The above LSDA metallic solution contradicts the experimental insulating
behavior. In order to probe the electron correlation effect, we now carry out
LSDA+$U$ calculations. The electron correlation stabilizes the Ni$^{2+}$ $S$=1
state and the calculated spin moment of Ni$^{2+}$ is enhanced to 1.68 $\mu_B$.
As seen in Fig. 3(a), the single Ir$^{4+}$ $t_{2g}$ hole now fully occupies the
down-spin $a_{1g}$ orbital. The Ir$^{4+}$ spin moment is also increased to
0.64 $\mu_B$. Owing to the correlation driven $d$-electron localization,
the induced $2p$ spin moment on oxygen via the Ir-O and Ni-O hybridizations
is reduced to 0.08 $\mu_B$. Apparently, electron correlations make the Ir $5d$
and Ni $3d$ spin-orbital states fully polarized, thus giving an insulating 
solution. The tiny band gap is within the Ir $t_{2g}$ shell, and it is due to
a relatively weak electron correlation of delocalized $5d$ electrons. 
As Ir$^{4+}$ has the single $t_{2g}$ hole on the down-spin 
$a_{1g}$ ($3z^2$--$r^2$) orbital, the down-spin $3z^2$--$r^2$ electron of 
Ni$^{2+}$ can hop forth and back. In order to maximize the local Hund exchange
on the virtual Ni$^{3+}$ ion in the excited intermediate state, the hopping
$3z^2$--$r^2$ electron should be in the minority-spin channel, see Fig. 4(b).
Then this exchange mechanism gives the FM coupling between Ir$^{4+}$ $S$=1/2
and Ni$^{2+}$ $S$=1. Indeed, our LSDA+$U$ calculation gives a total maximal 
spin moment of 3 $\mu_B$/fu for this FM insulating state. 
Here we note that the Ir$^{4+}$
$a_{1g}$ ($3z^2$--$r^2$) orbital is orthogonal to the Ni$^{2+}$ $xz$/$yz$ and
$xy$/$x^2$--$y^2$ orbitals, and hence there is no hopping between them to
account for the magnetism. Moreover, the Ir$^{4+}$ $e_g$ empty bands are too
high (see Fig. 2) to be relevant for the magnetic coupling. Therefore, with 
the crystal-field level diagrams depicted in Figs. 4(a) and 4(b), the electron
correlation and inter-orbital hybridization give rise to the FM insulating
solution.

As $5d$ transition metals have an intrinsic strong SOC and particularly iridates
are a representative example in this respect, now we are motivated to study
the SOC effect by doing LSDA+$U$+SOC calculations.  
It is interesting to note that now we get an AF insulating solution with 
a small band gap of 0.15 eV, see Fig. 3(b). Particularly, this solution has 
the Ir$^{4+}$ single $t_{2g}$ hole on the $e'_g$ orbital, 
in sharp contrast to the $a_{1g}$ 
hole state in the above LSDA+$U$ FM insulating solution. 
The Ni$^{2+}$ retains its configuration state and has a spin moment
of 1.69 $\mu_B$. Owing to the small crystal-field splitting of 0.19 eV 
between ($xy$, $x^2$--$y^2$) and ($xz$, $yz$), a finite mixing between them
due to the Ni SOC gives also a small orbital moment of 0.21 $\mu_B$ on Ni$^{2+}$.
The Ir$^{4+}$ ion has now a spin moment of --0.44 $\mu_B$. Moreover, the
$e'_g$ doublet can form a complex orbital with $l_z$ = $\pm$1. 
Then in the up-spin channel, the SOC
lowers $l_z$ = --1 state, and the $l_z$ = 1 state is pushed above the Fermi level
by SOC and moderate electron correlation, determining the modest band gap
and giving an orbital moment of --0.51 $\mu_B$ on Ir$^{4+}$. Owing to the 
significant Ir-O covalency, both the spin and orbital moments are reduced 
from their respective ideal unit value. 
In the AF state of Ni$^{2+}$ $S$=1 and Ir$^{4+}$
$S$=--1/2, the induced magnetic moment on oxygen gets tiny (only about 0.01
$\mu_B$).

The above results indicate an interesting evolution of the intrachain magnetic 
structure, from the LSDA+$U$ FM state to the LSDA+$U$+SOC AF state. It is 
ascribed to the SOC tuning orbital state of the
Ir$^{4+}$ ion. Although the $a_{1g}$ is a higher crystal-field level than the 
$e'_g$ by 0.21 eV, the significant SOC of the Ir$^{4+}$ ion (being 
about 0.5 eV) can well split the $e'_g$ doublet and eventually aligns the 
upper branch above the $a_{1g}$ (Fig. 4(c)). As a result, the single $t_{2g}$
hole of the Ir$^{4+}$ ion lies in the $e'_g$ state. Then, the up-spin 
$xz$/$yz$ electrons of the Ni$^{2+}$ ion can hop, forth and back, to
the up-spin $e'_g$ empty state (i.e., the up-spin $l_z$ = 1 branch), 
giving rise to the AF coupling via the
$dd\pi$ hybridization (Figs. 4(b) and 4(c)). Actually, using the orbital
states depicted in Figs. 4(b) and 4(c), we also calculated the FM state. 
Our LSDA+$U$+SOC calculations find that the FM state is indeed less stable
than the AF ground state by 110 meV/fu. As the SOC is intrinsic in iridates,
the AF ground state is deemed reliable from the LSDA+$U$+SOC calculations,
but the FM state seems fictitious from the LSDA+$U$ calculations 
without inclusion of the SOC. Indeed, the AF ground state agrees with the 
most recent experiment~\cite{Tjeng}. Therefore, we can conclude that
it is the significant SOC of the Ir$^{4+}$ ion which tunes the spin-orbital
states and Ir-Ni inter-orbital interactions and hence determines the 
AF structure of {\SNIO}.

\begin{figure}[h]
\centering \includegraphics[width=8cm]{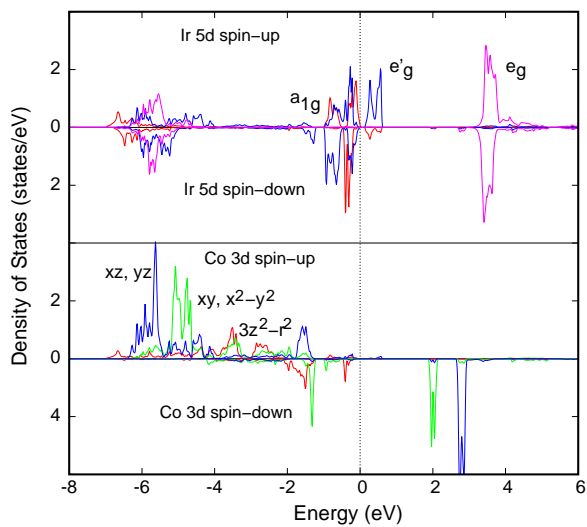} 
\caption{(Color online)
The partial DOS of the Ir $5d$ and Co $3d$ eigen orbitals
in AF {\SCIO} calculated by LSDA+$U$+SOC. The high-spin Co$^{2+}$ ion ($S$ = 3/2)
has the orbitals $3z^2$--$r^2$ and [($x^2$--$y^2$) + i$xy$]/$\sqrt{2}$ ($l_z$ = 2)
occupied in its down-spin channel. The Ir$^{4+}$ ($S$ = --1/2) has the $e'_g$ 
$l_z$ = 1 empty state in its up-spin channel.
}
\end{figure}

Now we turn to {\SCIO}. As this material has practically the same 
crystal structure as {\SNIO}~\cite{Nguyen,Tjeng},
both systems have many common features
in the electronic and magnetic structures.
The $a_{1g}$ singlet of the Ir$^{4+}$ ion is higher than the $e'_g$ doublet in 
the crystal-field level diagram. Moreover, the high-spin Co$^{2+}$ ion has the 
same crystal-field level sequence as Ni$^{2+}$, but it now has one hole
on the $xy$/$x^2$--$y^2$ doublet (compared with Ni$^{2+}$, see Fig. 4(b)).
Our LSDA+$U$ calculations give a FM metallic solution due to the $3z^2$--$r^2$
electron hopping and the 3/4 filled $xy$/$x^2$--$y^2$ bands. 
Apparently, this solution contradicts the experimental AF insulating 
behavior~\cite{Tjeng}.

However, when we include SOC by doing LSDA+$U$+SOC calculations, we have
obtained the correct AF insulating solution (see Fig. 5) in good agreement
with the experiment~\cite{Tjeng}. 
The high-spin Co$^{2+}$ ion ($S$=3/2) has a spin moment of
2.66 $\mu_B$. In its down-spin channel, the $xy$/$x^2$--$y^2$ doublet 
form the complex orbitals ($x^2$--$y^2$) $\pm$ i$xy$ with $l_z$ = $\pm$2, 
and the Co$^{2+}$ SOC lowers the $l_z$ = 2 
state but lifts the $l_z$ = --2 state. The electron correlation places 
the former at --1.5 eV and the latter at 2 eV. As a result, the Co$^{2+}$ ion
has also a huge orbital moment of 1.71 $\mu_B$. In total, the Co$^{2+}$ ion 
has the magnetic moment of 4.37 $\mu_B$ and it is firmly aligned, due to the
SOC, along the hexagonal $c$-axis (i.e., a significant Ising-like 
spin system). Moreover, the Ir $5d$ states are almost the same as in {\SNIO}:
the Ir$^{4+}$ SOC places the single $t_{2g}$ hole on the $e'_g$ doublet; 
the SOC and moderate electron correlation determine the small insulating gap
within the Ir$^{4+}$ $t_{2g}$ shell, see Fig. 5. 
The Ir$^{4+}$ ion has the spin (orbital) moment of --0.39 (--0.47) $\mu_B$ and 
in total --0.86 $\mu_B$. Using these spin-orbital states, our LSDA+$U$+SOC
calculations find that the AF ground state is more stable than the FM state
by 122 meV/fu.

By looking at Figs. 4(b) and 4(c), and now having also one hole on the 
down-spin $xy$/$x^2$--$y^2$ doublet for Co$^{2+}$, we find that the net spin=1
from $xz$/$yz$ and the net spin=1/2 from $xy$/$x^2$--$y^2$ both 
contribute to the AF exchange with the net Ir spin = --1/2 from the $e'_g$.
The former is via a $dd\pi$ hybridization as in {\SNIO}, and the latter a weaker 
$dd\delta$ one (which is missing in {\SNIO}). 
This accounts for a relatively higher stability of 
the AF ground state over the FM state in {\SCIO} than in {\SNIO}, i.e., 122 $vs$
110 meV/fu. In addition, the $dd\delta$ hybridization results in the 
smaller spin and orbital moments of the Ir$^{4+}$ ion in {\SCIO} than in {\SNIO}.
Note that all these results qualitatively explain the slightly higher intrachain
AF transition temperature of 90 K in {\SCIO} than 85 K in {\SNIO}~\cite{Tjeng}.
Moreover, the magnetic moments of the significant Ising magnetism, 4.37 $\mu_B$/Co$^{2+}$ 
and --0.86 $\mu_B$/Ir$^{4+}$, also
agree reasonably well with the experimental ones of 3.6 and 
--0.6 $\mu_B$~\cite{Tjeng}.


In summary, using density functional calculations including SOC 
and electron correlation, we have demonstrated that the hexagonal spin-chain 
materials {\SMIO} ($M$=Ni,Co) are another iridate system in which the SOC tunes 
the magnetic and electronic properties. The significant SOC alters the orbital 
state, the exchange pathway, and thus the magnetic structure. 
We therefore have clarified the nature and the origin of the intrachain AF structure. 
The present results well account for the most recent experiments.\\

This work was supported by the NSF of China (Grant No. 11274070), Pujiang Program
of Shanghai (Grant No. 12PJ1401000), and 
ShuGuang Program of Shanghai (Grant No. 12SG06).

\end{document}